# Correlation-driven eightfold magnetic anisotropy in a two-dimensional oxide monolayer

Zhangzhang Cui,[1,2] Alexander J. Grutter,[3] Hua Zhou,[4] Hui Cao,[1,2,4] Yongqi Dong,[1,4] Dustin A. Gilbert,[3,5] Jingyuan Wang,[6] Yi-Sheng Liu,[7] Jiaji Ma,[8] Zhenpeng Hu,[9] Jinghua Guo,[7] Jing Xia,[6] Brian J. Kirby,[3] Padraic Shafer,[7] Elke Arenholz,[7,10] Hanghui Chen,[8,11,12]* Xiaofang Zhai,[1,2]* Yalin Lu[1,2]

[1]Hefei National Laboratory for Physical Sciences at Microscale, National Synchrotron Radiation Laboratory, University of Science and Technology of China, Hefei, Anhui 230026, China

[2]Synergetic Innovation Center of Quantum Information and Quantum Physics, University of Science and Technology of China, Hefei, Anhui 230026, China

[3]NIST Center for Neutron Research, National Institute of Standards and Technology, Gaithersburg, Maryland 20899, USA

[4]Advanced Photon Source, Argonne National Laboratory, Lemont, Illinois 60439, USA

[5]Department of Materials Science and Engineering, University of Tennessee, Knoxville, TN 37996, USA

[6]Department of Physics, University of California at Irvine, Irvine, California 92697, USA

[7]Advanced Light Source, Lawrence Berkeley National Laboratory, Berkeley, California 94720, USA

[8]NYU-ECNU Institute of Physics, NYU Shanghai, Shanghai 200122, China

[9]School of Physics, Nankai University, Tianjin 300071, China

[10]Cornell High Energy Synchrotron Source, Cornell University, Ithaca, New York 14853, USA

[11]State Key Laboratory of Precision Spectroscopy, School of Physical and Material Sciences, East China Normal University, Shanghai 200062, China

[12]Department of Physics, New York University, New York 10027, USA

[13]School of Physical Science and Technology, ShanghaiTech University, Shanghai 201210, China.

*Corresponding emails: zhaixf@shanghaitech.edu.cn (X.Z.) or hanghui.chen@nyu.edu (H.Ch.)

**Engineering magnetic anisotropy in two-dimensional systems has enormous scientific and technological implications. The uniaxial anisotropy universally exhibited by two-dimensional magnets has only two stable spin directions, demanding 180° spin switching between states. We demonstrate a novel eightfold anisotropy in magnetic $SrRuO_3$ monolayers by inducing a spin reorientation in $(SrRuO_3)_1/(SrTiO_3)_N$ superlattices, in which the magnetic easy axis of Ru spins is transformed from uniaxial $\langle 001 \rangle$ direction ($N < 3$) to eightfold $\langle 111 \rangle$ directions ($N \geqslant 3$). This eightfold anisotropy enables 71° and 109° spin switching in $SrRuO_3$ monolayers, analogous to 71° and 109° polarization switching in ferroelectric $BiFeO_3$. First-principle calculations reveal that increasing the $SrTiO_3$ layer thickness induces an emergent correlation-driven orbital ordering, tuning spin-orbit interactions and reorienting the $SrRuO_3$ monolayer easy axis. Our work demonstrates that correlation effects can be exploited to substantially change spin-orbit interactions, stabilizing unprecedented properties in two-dimensional magnets and opening rich opportunities for low-power, multi-state device applications.**

**INTRODUCTION**

Recent years have seen intense interest in stabilizing and controlling magnetic ordering in two-dimensional (2D) systems (*1-9*), motivated by both the potential to unlock new fundamental physics and enable new high-density, low-power spintronic device paradigms. Engineering magnetic anisotropy (MA) in 2D systems plays a critical role in realizing these new functionalities, but remains challenging due to the lack of accessible control parameters. Atomically precise complex oxide superlattices provide an ideal platform for the manipulation of MA in magnetic monolayers, as the correlated electron physics enables uniquely powerful handles through strong coupling between the electronic, spin, orbital, and lattice degrees of freedom. These handles allow deterministic control of the electronic and magnetic ground state, leading to exotic phenomena such as high-temperature superconductivity, colossal magnetoresistance, 2D electron gases, etc. (*10,11*)

High-quality $(SrRuO_3)_1/(SrTiO_3)_N$ superlattices, in which each $SrRuO_3$ monolayer is separated by $N$ unit cells of $SrTiO_3$, are an ideal model system in which to explore the interplay between electron correlation and MA. The MA of a $SrRuO_3$ monolayer originates from strong spin-orbit interactions. Atomic spin-orbit coupling (SOC) is proportional to $Z^4$ (where $Z$ is atomic number) (*12,13*), so that a 4*d* transition metal such as Ru exhibits a larger SOC energy (about 100 meV) than 3*d* transition metals (*14*). $Ru^{4+}$ in bulk $SrRuO_3$ nominally has four *d*-orbital electrons in a low-spin configuration, where three electrons occupy the majority spin channel while the fourth electron resides in the minority spin channel with occupational degeneracy among the three Ru $t_{2g}$ orbitals (*15-18*). In this work, we propose to use oxide superlattices to tune Ru orbital occupancy, which changes the SOC energy and induces a nontrivial new MA in $SrRuO_3$ monolayers.

**RESULTS**

**Structural characterizations of $(SrRuO_3)_1/(SrTiO_3)_N$ superlattices**

The $(SrRuO_3)_1/(SrTiO_3)_N$ superlattices are shown schematically in Fig. 1A and were fabricated by pulsed laser deposition (PLD) assisted with reflective high energy electron diffraction (RHEED). X-ray diffraction (XRD) measurements in Fig. 1B reveal superlattice peaks corresponding to the designed periodicity. Layer-by-layer growth and atomically flat surfaces are observed by in-situ RHEED and atomic force microscopy respectively (Fig. S1). The x-ray absorption near-edge structure (XANES) of Ru *K*-edges are measured which demonstrate similar Ru valences in the superlattices (Fig. S2). XRD reciprocal space maps around the (2 0 4) substrate peak are shown in Fig. 1C, demonstrating that all superlattices are coherently strained to the $SrTiO_3$ substrates. The average *z*-axis lattice constants $c_{average}$ are calibrated and shown in Fig. 1D. The ideal *z*-axis lattice constants calculated as $c_{ideal} = (N \times c_{STO} + c_{SRO})/(N+1)$ are used to fit $c_{average}$, where $c_{STO}$ and $c_{SRO}$ represent that of $SrTiO_3$ (3.905 Å) and $SrRuO_3$ (3.984 Å) respectively. This comparison shows that, within experimental uncertainty, $c_{average}$ matches $c_{ideal}$ across all $N$, so that the lattice constants and strain states of all superlattices are consistent.

Furthermore, we measured half order diffraction peaks to reveal the oxygen octahedral rotation patterns (*19-21*). In all superlattices $a^-a^-c^-$ rotation patterns are observed (see section SI in the Supplementary Materials for details). Fig. 1E shows the (3/2 1/2 3/2) and (3/2 1/2 5/2) half order peaks with stronger intensities in superlattices of smaller $N$. Additionally, extremely weak (H/2, H/2, L/2) diffraction peaks are observed in all superlattices (Fig. S3, A and B), indicating that the residual $a^-$ rotation is much smaller than the $c^-$ rotation. We therefore conclude that all superlattices exhibit tetragonal structural symmetry with $a^-a^-c^-$ type octahedral rotations, where the $c^-$ rotations are larger than the $a^-$ rotations.

**Magnetism and Curie temperatures of (SrRuO$_3$)$_1$/(SrTiO$_3$)$_N$ superlattices**

In Fig. 2A and 2B, we show the Ti *L*-edge x-ray absorption spectroscopy (XAS) and x-ray magnetic circular dichroism (XMCD) of the superlattices measured at a temperature (*T*) of 10 K in an applied magnetic field (*H*) of 4 T. As the absorption energy of the Ru *M*- and Ti *L*-edges overlap, the XAS and XMCD are completely dominated by Ti so that it is not possible to distinguish the Ru *M*-edge signal in the superlattices (*22*). More XMCD data with both normal and grazing incidence beam of the other superlattices are shown in Supplementary Materials Fig. S4. No measurable valence change or magnetic dichroism was observed on the Ti edge in all superlattices, excluding any magnetic contribution from the SrTiO$_3$ and indicating that the magnetization (*M*) is confined purely within the SrRuO$_3$ layers. The magnetism and MA of the SrRuO$_3$ monolayer are further revealed by superconducting quantum interference device (SQUID) magnetometer and magneto-optic Kerr effect (MOKE) measurements. The former detects the overall magnetism from the film and possible artificial backgrounds, while the latter only detects the film with an optical penetration depth (~30 nm) less than the film thickness. Note that the cooling fields to orient the magnetic domains are 0.05 T and 0.5 T for the MOKE and SQUID measurements, respectively. The lower cooling field yielded the low temperature peak features in some of the MOKE measurements. The temperature-dependent Kerr rotation (Fig. 2C, right axis and Fig. S5A in the Supplementary Materials) reveals Curie temperatures ($T_C$) of approximately 100 K for the *N* = 1 and 70 K for the *N* = 2-5 superlattices. Thus the magnetic transition can be confirmed to be intrinsic to the films.

The quantitative magnetizations of the superlattices are studied by SQUID *M-H* measurements (Fig. 2, D-F and Fig. S5, B and C), which reveals a saturation magnetization between 0.5 $\mu_B$/Ru and 0.7 $\mu_B$/Ru for the *N* = 1 and 2 superlattices and approximately 0.4 $\mu_B$/Ru for the *N* ≥ 3 superlattices. Strikingly, the MA of the superlattices exhibits a significant dependence on the SrTiO$_3$ layer number. The *N* ≤ 2 superlattices are similar to the bulk, remaining uniaxial with the easy axis along [001] (see Fig. 2D and Fig. S5B). However, the magnetic hysteresis of the *N* ≥ 3 superlattices indicates an easy axis transition to be along the [111] direction (see Fig. 2, E and F and Fig. S5C).

**Depth-dependent Magnetization Distribution**

Since the Ru *M*-edge XMCD was not detectable, we have probed the magnetization distribution in the *N* = 3 superlattice with polarized neutron reflectometry (PNR), as shown in Fig. 3A. Although the superlattice repeat length is extremely thin, so that the first order Bragg reflection appears at approximately 4 nm$^{-1}$, the extremely sharp interfaces and sample uniformity allow the observation of a clear superlattice peak in the expected location, as shown in Fig. 3B. Because the neutron spin provides sensitivity to the magnetic scattering length densities, analysis and model fitting of the PNR data allows a depth-dependent picture of the magnetization distribution to be extracted, shown in Fig. 3D. Specifically, we note that a nonzero splitting (see Fig. 3C) is observed between the (++) and (--) reflectivities near the critical edge, which indicates an intrinsic net magnetization within the film of at least 0.24 $\mu_B$/Ru and up to 0.37 $\mu_B$/Ru. The PNR detected magnetization is slightly smaller than the SQUID magnetization, but agrees reasonably well given possible background contributions to the SQUID value. Further, a small but statistically notable spin asymmetry (SA), defined as $(R^{++} - R^{--})/(R^{++} + R^{--})$ of 0.167 ± 0.045 was observed at the first-order Bragg reflection. Modeling indicates that the SA of this feature is highly dependent on which layer the net magnetization originates in, with magnetic SrRuO$_3$ yielding a positive SA and magnetic SrTiO$_3$ yielding a negative SA. Since the observed SA is clearly positive, we conclude with high confidence that the observed magnetism originates from the SrRuO$_3$ layers as expected. Model fitting of the data supports this interpretation, with an approximate fitted magnetic moment of 0.004 $\mu_B$/Ti ± 0.055 $\mu_B$/Ti in the SrTiO$_3$ layers. We therefore conclude that PNR reveals net magnetization originating from the SrRuO$_3$ monolayers in excellent agreement with the SQUID, MOKE and Ti XMCD measurements.

**MA of (SrRuO$_3$)$_1$/(SrTiO$_3$)$_N$ superlattices**

To reveal the exact symmetry of the MA, we perform transverse magnetoresistance (MR) and magnetic-field angle-dependent resistance (MAR) measurements. The MR was measured at 5 K with the current driven along the [100] direction (Fig. 4, A and B). The MR of *N* = 1, 2 superlattices with magnetic field *H* // [001] shows a two-peak structure with lobes reflecting the magnetic hysteresis loops. In contrast, the hysteresis loops are suppressed in the MR with *H* // [010], consistent with the weaker in-plane magnetization of the *N* = 1, 2 superlattices. The MR measurements of the *N* = 3-5 superlattices all show similar behavior in the *H* // [001] and *H* // [010] measurements, indicating symmetric in-plane and out-of-plane spin alignments. Fig. 4C presents the polar plots of the MAR of (SrRuO$_3$)$_1$/(SrTiO$_3$)$_N$ superlattices measured at *H* = 9 T and at temperatures of 5 K, 25 K and 50 K. The MAR of *N* = 3 and 5 superlattices with 5 K temperature steps are shown in Fig. S6, A and B. Here we define MAR as MAR = $(\rho(\theta) - \rho(90°))/\rho(90°)$, where $\rho$ is the resistivity and $\theta$ represents the angle between the magnetic field *H* and the film surface normal (see the inset of Fig. 4B). *H* was rotated in the (100) plane with the electric current maintained perpendicular to the field. The *N* = 1, 2 samples exhibit perpendicular MA at all measured temperatures, as do all other samples at *T* > 25 K, consistent with the perpendicular MA identified in the SQUID measurements. For *N* ≥ 3, we observe a transition from twofold perpendicular MA to fourfold MA with decreasing temperature. The magnetic easy axes at low temperatures are along the [011], [0$\bar{1}$1], [01$\bar{1}$] and [0$\bar{1}\bar{1}$] directions. The MAR at 5 K with *H* rotating in the (010) and (001) planes is similar with that of (100) plane (Fig. S6, C and D), as expected given the pseudo-cubic structure of SrRuO$_3$ crystal lattice in the superlattices. More comprehensive MAR measurements at 5 K of the $N$ = 3 superlattice with $H$ rotating in the (110) plane (Fig. 4D) reveal an angle of ~71º/109º between the two magnetic easy axes within the (110) plane. Thus, the observed MAR symmetry identifies magnetic easy axes along the eightfold ⟨111⟩ directions of the SrRuO$_3$ pseudo-cubic lattice. These results confirm that perpendicular MA exists in SrRuO$_3$ monolayers for $N ≤ 2$ superlattices at all temperatures and in $N ≥ 3$ superlattices above ~25 K. Below approximately 25 K (±5 K), the SrRuO$_3$ monolayers in $N ≥ 3$ superlattices exhibits eightfold MA.

**First-principle calculations of the MA**

In order to understand why MA of (SrRuO$_3$)$_1$/(SrTiO$_3$)$_N$ superlattices changes with the thickness of SrTiO$_3$ at low temperatures, we perform first-principle calculations. The in-plane lattice constants (along $x$ and $y$ axes) of all superlattices are constrained to match the theoretical lattice constant of the SrTiO$_3$ substrate. We start from a crystal structure with the experimentally observed $a^-a^-c^-$ rotation pattern (space group No. 14, $P2_1/c$). After atomic relaxation, density functional theory (DFT) calculations find a large rotation angle $\gamma$ about the $z$-axis but a very small rotation angle $\alpha$ about $x$-axis and $y$-axis ($\approx 0.5$º) in both $N$ = 1 and $N$ = 3 superlattices, consistent with the XRD results. The layer-resolved rotation angles $\alpha$ and $\gamma$ of each oxygen octahedron are shown in Fig. 5A and Fig. 5D. We note that the calculated $\gamma$ angles from our calculations are very similar to those reported in a previous study (*5*).

Fig. 5B shows the near-Fermi-level density of states (DOS) of the $N$ = 1 superlattice. Ru in SrRuO$_3$ has a formal $d^4$ occupancy, with three electrons occupying the majority spin state (upper half of the panel) and the fourth electron in the minority spin state (lower half of the panel). The SrRuO$_3$/SrTiO$_3$ interfaces remove the degeneracy between Ru $d_{xy}$ and $d_{xz/yz}$, so that the fourth (minority spin) electron is evenly shared by Ru $d_{xz}$ and Ru $d_{yz}$ orbitals. This electronic structure is consistent with previous results (*5,15*). Turning on spin-orbit coupling to induce MA, we test three different magnetic moment orientations: along ⟨001⟩, ⟨100⟩ and ⟨111⟩ directions. We find that in the $N$ = 1 superlattice, the ⟨001⟩ state has the lowest total energy among the three magnetic orientations (Fig. 5C), in agreement with the SQUID and magnetotransport measurements. The twofold ⟨001⟩ MA is explicitly shown in the inset of Fig. 5C.

However, in the $N$ = 3 superlattice, we find a completely new correlated state with different electronic, magnetic and orbital properties. Fig. 5E shows the near-Fermi-level DOS of the $N$ = 3 superlattice, which indicates semiconducting behavior with a small band gap of about 0.1 eV, in agreement with the transport measurements (Fig. S7A). More importantly, in the $N$ = 3 superlattice, in the minority spin channel, Ru $d_{xz}$ and Ru $d_{yz}$ orbitals hybridize into a pair of new orbitals Ru $\alpha|xz\rangle+\beta|yz\rangle$ orbital (referred to as Ru (+) state) and Ru $\beta|xz\rangle-\alpha|yz\rangle$ orbital (referred to as Ru (-) state), where $\alpha^2+\beta^2=1$. From our DFT+$U$ calculations, we find $\alpha \sim \beta \sim 1/\sqrt{2}$. In each RuO$_2$ plane, there are two distinct Ru atoms: on one Ru atom, the fourth electron fills Ru (+) state and leaves Ru (-) state empty; on the other Ru atom, the fourth electron fills Ru (-) state and leaves Ru (+) state empty. The filled new orbital is referred to as a lower Hubbard band, which is just below the Fermi level; the empty new orbital is referred to as an upper Hubbard band, which is about 2 eV above the Fermi level. Such an orbital ordering is very similar to what is found in layered perovskite K$_2$CuF$_4$, in which the hole orbitals $|x^2-r^2\rangle$ and $|y^2-r^2\rangle$ alternate in a basal plane (*23*). This orbital ordering results in a ferromagnetic insulating state in the CuF$_2$ plane according to Goodenough-Kanamori-Anderson rule (*24-26*). The emergence of the new orbital ordering in the $N$ = 3 superlattice is corroborated with the fact that in each RuO$_2$ layer, Ru has one pair of long Ru-O bonds and one pair of short Ru-O bonds (2.06 Å and 1.97 Å, respectively) in our DFT calculation. Such a bond disproportionation has also been observed in our calculated results of $N$ = 5 superlattice and in K$_2$CuF$_4$ (*23*). On the other hand, in the $N$ = 1 superlattice in which the new orbital ordering does not occur, our calculation shows that Ru has four equal Ru-O bonds in the RuO$_2$ plane (2.00 Å).

It is precisely this new orbital ordering that changes MA. To demonstrate this, we turn on SOC and find that in the $N$ = 3 superlattice the ⟨001⟩ state does not have the lowest energy, but rather the ⟨111⟩ state becomes the most stable among the three magnetic orientations considered (Fig. 5F), which is consistent with the key experimental discovery as described above. The eightfold ⟨111⟩ MA is explicitly shown in the inset of Fig. 5F. The DFT calculation of $N$ = 5 superlattice is similar to that of $N$ = 3 and the results are shown in Fig. S8 in the Supplementary Materials. The reason a new correlated state emerges in the $N$ = 3 and 5 superlattices is that with the RuO$_2$ layers further separated, inter-planar Ru-Ru hopping is suppressed, decreasing the band width of Ru anti-bonding states (Fig. 5, B and E) and increasing correlation effects on the Ru sites. Furthermore, the rotations of oxygen octahedra reduce the crystal symmetry, contributing to the removal of the orbital degeneracy (Fig. S9). The two factors combined lead to a hybridization of Ru $d_{xz}$ and Ru $d_{yz}$ orbitals and a split into a pair of lower and upper Hubbard bands. The role of oxygen octahedral tilts on the electronic structure is discussed in section SII in the Supplementary Materials. The new correlation-driven orbital ordering and the resulting eightfold ⟨111⟩ MA of (SrRuO$_3$)$_1$/(SrTiO$_3$)$_N$ ($N ≥ 3$) superlattices are different from those of magnetic interfaces in previous studies (*27-31*).

**DISCUSSION**

Our study reveals a novel eightfold ⟨111⟩ MA in SrRuO$_3$ monolayers in (SrRuO$_3$)$_1$/(SrTiO$_3$)$_N$ superlattices ($N \geq 3$). Theoretically, our first-principle calculations demonstrate that the enhanced correlation strength on Ru atoms leads to a metal-to-semiconductor transition and induces an orbital ordering that is different from that of $N = 1$ superlattice, but is similar to ferromagnetic insulator K$_2$CuF$_4$. The emergent orbital ordering changes the underlying spin-orbit interaction, reorienting the Ru magnetic easy axis. Experimentally, we performed four independent measurements (SQUID, MOKE, PNR, and MR) to understand the magnetic property of (SrRuO$_3$)$_1$/(SrTiO$_3$)$_N$ superlattices. First we find that paramagnetism is unlikely because we observe hysteresis loops in both SQUID and transverse MR measurements, which is the characteristic feature of ferromagnetic materials. In addition, the temperature dependence of the Kerr rotation is incompatible with the usual Curie-Weiss behavior of paramagnetism. Second, the saturation magnetization around 0.3 $\mu_B$/Ru in $N \geq 3$ superlattices is much larger than the usual net moment of canted antiferromagnetism in complex oxides (*32-35*). Furthermore, the reasonable agreement between the saturation magnetizations measured by PNR and SQUID does not support the possibility of an unintentionally subtracted linear *M*-versus-*H* dependence which is the fingerprint proof of canted antiferromagnetism (*35-38*). On the other hand, ferromagnetism with a relatively small saturation moment is compatible with all the results we have obtained and we consider it as the most likely magnetic property in the large *N* superlattices.

Our work demonstrates that tuning interlayer electron hopping via digital oxide superlattices is a powerful tool for controlling spin-orbit interaction in solids and inducing novel physical properties in 2D magnetic monolayers, which are not exhibited by their bulk counterparts. MA with symmetry higher than fourfold is extremely rare in bulk magnetic materials, let alone 2D magnetic monolayers. The new eightfold ⟨111⟩ MA in a magnetic monolayer has far-reaching scientific and technological implications, such as multi-state memory devices with eight degenerate magnetic states in real space, spin transfer torque or spin orbit torque with a minimum of 71° spin switching (which will substantially reduce the critical current), and control of topological spin textures when inversion symmetry is broken (*39-41*).

**MATERIALS AND METHODS**
**Preparation and structural characterizations of $(SrRuO_3)_1/(SrTiO_3)_N$ superlattices**
The $(SrRuO_3)_1/(SrTiO_3)_N$ ($N$ = 1-5) superlattices were fabricated on (001) $SrTiO_3$ substrates using single-crystalline $SrTiO_3$ and ceramic $SrRuO_3$ targets by PLD assisted with RHEED. $SrTiO_3$ substrates with atomically flat $TiO_2$ termination were obtained via buffered Hydrofluoric acid etching and annealing. The RHEED system was used to monitor the layer-by-layer growth of the films and the total repetitions of the $N$ = 1 to 5 superlattices are all 50. The thicknesses of both the $SrRuO_3$ layer and the $SrTiO_3$ layer are precisely controlled at a single molecular level by RHEED. All films were grown at a substrate temperature around 700 °C and under an oxygen pressure of 10 Pa. During the growth, the laser frequency and energy density were 2 Hz and ~1 J/cm$^2$, respectively. After the deposition, all films were in situ annealed at 500 °C for an hour in an oxygen environment $5\times10^4$ Pa to remove oxygen vacancies. Synchrotron XRD measurements were conducted at the Advanced Photon Source, Argonne National Laboratory on beamline 12-ID-D using the Pilatus 100K detector, and at the Shanghai Synchrotron Radiation Facility on beamline 14B.

**Magnetic and magnetotransport characterizations**
The magnetic properties of the superlattices were probed using SQUID and MOKE techniques. The temperature-dependent magnetization and Kerr rotation measurements of the superlattices were done during warming up under a smaller field of 0.05 T after the samples were first cooled down to 4 K under a field of 0.5 T for SQUID and 0.05 T for MOKE. The transport properties were measured using a standard linear four-probe method by a Physical Property Measurement System (PPMS) equipped with a sample rotator. Au electrodes were deposited using Ar ion sputtering on top of the superlattices. During the transport measurements, the DC current of around 10 µA was applied to the films and the direction of the current was maintained to be perpendicular to the magnetic field.

**Polarized Neutron Reflectometry measurements**
PNR measurements were performed using the polarized beam reflectometer instrument at the NIST Center for Neutron Research. Samples were cooled to 6 K in an applied field of 3 T. Full polarization analysis was performed using both a spin-polarizer and spin-analyzer. The spin-dependent reflectivity was measured as a function of the scattering vector Q along the film normal. Data was reduced with the Reductus software package (*42*) and analyzed with the Refl1D software package for reflectometry modeling (*43*). Uncertainties in fitted parameters were extracted using a Markov chain Monte-Carlo algorithm Differential Evolution Adaptive Metropolis (DREAM) as implemented in the BUMPS python package. We note that polarized neutron reflectometry is sensitive only to the net in-plane components of the magnetization within the film, so that any out-of-plane component, for example, canted towards ⟨111⟩ axes will not be observed. For that reason, the reported magnetization values have been adjusted to account for the fact that SQUID magnetometry indicates the films are approximately 10% below saturation value at 3 T. We also note that since no in-plane perpendicular magnetization component is expected in an applied field of 3 T, the spin-flip reflectivities $R^{+-}$ and $R^{-+}$ are expected to be zero and were not collected. Only the non-spin-flip scattering cross sections $R^{++}$ and $R^{--}$ were measured.

**X-ray spectroscopic measurements**
The Ti $L$-edge XAS and XMCD measurements were performed on beamline 4.0.2 at the Advanced Light Source (ALS) of Lawrence Berkeley National Laboratory (LBNL) at a temperature of 10 K and under the vacuum pressure of ($\approx$ or <) $1\times10^{-6}$ Pa. The XAS spectra were recorded in total electron yield mode (TEY, sample-to-ground drain current) and normalized by the incident photon flux determined from the photocurrent of an upstream Au mesh. The samples were measured with alternating left-polarized ($\mu^+$) and right-polarized ($\mu^-$) photons at 10 K cooled by liquid helium in an applied field of 4 T. During the XMCD measurement, the incident beam was perpendicular or inclined with a grazing angle of 20° to the sample surface and the spectra were collected in both TEY and luminescence yield mode. Preliminary room-temperature XAS measurements have been performed on beamline 8.0.1 at ALS, beamline BL12B-a at the National Synchrotron Radiation Laboratory of China. The XANES measurements at Ru $K$-edge were performed at the beamline 12-BM-B, and the x-ray linear dichroism measurements at Ru $L_3$-edge were carried out at the beamline 4-ID-D at the Advanced Photon Source of Argonne National Laboratory.

**First-principle calculations**
We perform DFT calculations using a plane wave basis set and projector-augmented wave method (*44*), as implemented in the Vienna Ab-initio Simulation Package (VASP) (*45*). We use Perdew-Burke-Ernzerhof (PBE) generalized gradient approximation as exchange correlation functional (*46*). An energy cutoff of 600 eV is used throughout the calculations. The Brillouin zone integration is performed with a Gaussian smearing of 0.05 eV over a Γ-centered $k$-mesh of $12 \times 12 \times 12$ for the $N$ = 1 superlattice and a Γ-centered $k$-mesh of $12 \times 12 \times 6$ for the $N$ = 3 and 5 superlattices. The threshold of self-consistent calculations is $10^{-6}$ eV. Crystal structure is relaxed until each force component is smaller than 0.01 eV/Å. The in-plane lattice constant is fixed to be 3.93 Å, which is the theoretical lattice constant of $SrTiO_3$ calculated by DFT-PBE method. Correlation effects on Ru atoms are taken into account (*47*) by using the rotationally invariant Hubbard *U* method in DFT calculations (DFT+*U* method) (*48*). Following the previous study, we employ $U_{Ru}$ = 4 eV (*5*). The key results do not qualitatively change for $U_{Ru} \geq 3$ eV. SOC is turned on to study magnetic anisotropy in DFT+*U*+SOC calculations.

**Acknowledgments**: The authors thank Prof. Darrell Schlom, Prof. Yuri Suzuki, Prof. Andrew Millis, Prof. Lu Li, Prof. Di Yi for helpful discussions, and Prof. Chung-Li Dong and Dr. Yongseong Choi for extensive beamline supports. **Funding:** This work was supported by the National Key Research and Development Program of China (Grants No. 2016YFA0401004), National Natural Science Foundation of China (Grants No. 51627901 and No. 11574287). The Advanced Light Source is supported by the Director, Office of Science, Office of Basic Energy Sciences, of the U.S. Department of Energy under Contract No. DE-AC02-05CH11231. The Advanced Photon Source, a U.S. Department of Energy Office of Science User Facility, is operated by Argonne National Laboratory under Contract No. DE-AC02-06CH11357. Sagnac measurement at UC Irvine was supported by NSF grant DMR-180781. X.Z. acknowledges the support of the Youth Innovation Promotion Association CAS (Grant No. 2016389). H.H.C. acknowledges the funding of National Natural Science Foundation of China (Grant No. 11774236), Shanghai Pujiang Talents Program (Grant No. 17PJ1407300), Seed grant of NYU-ECNU Research Institute of Physics and NYU University Research Challenge Fund. Computational resources are provided by NYU HPC resources at New York campus, Abu Dhabi and Shanghai. Certain commercial equipment is identified in this paper to foster understanding. Such identification does not imply recommendation or endorsement by the National Institute of Standards and Technology, nor does it imply that the materials or equipment identified are necessarily the best available for the purpose. **Author contributions:** X.Z and Z.C. designed the project. Y.L. supervised the work done in USTC. Z.C. fabricated the samples, performed the magnetic and transport measurements. H.Ch. and J.M. performed the theoretical calculations. A.J.G., B.J.K., D.A.G., P.S. and E.A. performed the low temperature XAS, XMCD and PNR measurements. H.C., Y.D. and H.Z. performed the synchrotron XRD, Ru $K$-edge XANES and Ru $L_3$-edge x-ray linear dichroism (XLD) experiments. J.W and J.X performed the MOKE measurement. X.Z., H.Ch., Z.C. and A.J.G. wrote the paper. All authors discussed the experimental data and commented to the manuscript writing. **Competing interests:** The authors declare that they have no competing interests. **Data and materials availability:** All data needed to evaluate the conclusions in the paper are present in the paper and/or the Supplementary Materials. Additional data related to this paper may be requested from the authors.

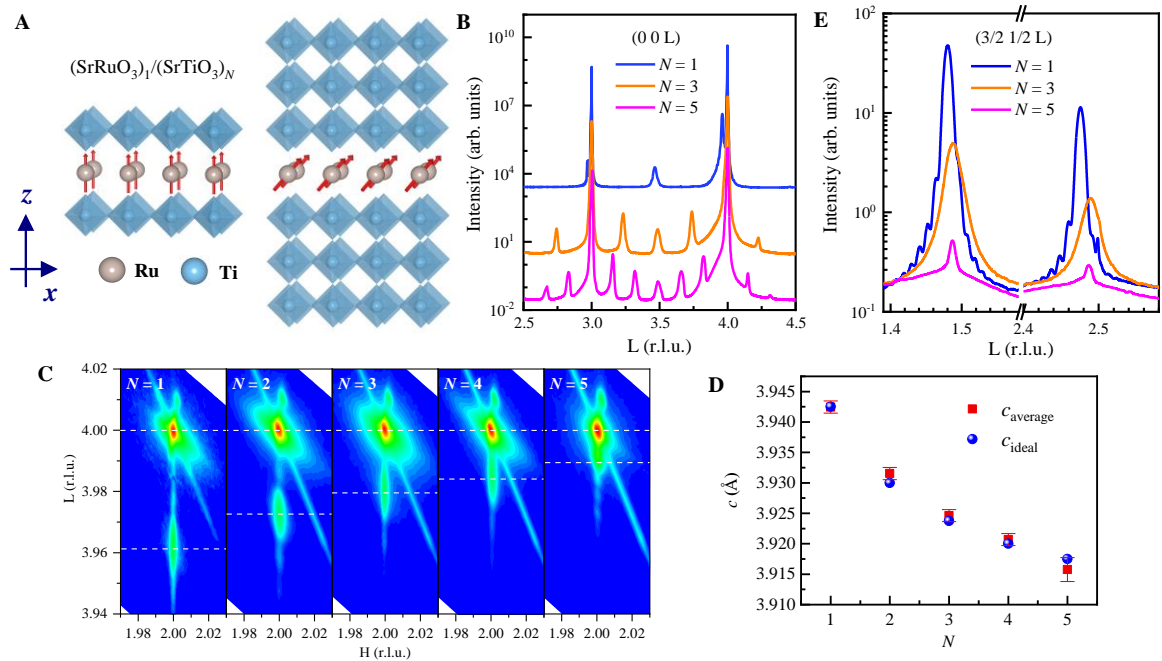

**Fig. 1. Structural characterizations of $(SrRuO_3)_1/(SrTiO_3)_N$ superlattices.** (**A**) Schematics of lattice structures of the $N = 1$ and $N = 3$ superlattices. (**B**) XRD $\omega$-$2\theta$ scans of $N = 1, 3, 5$ superlattices. (**C**) XRD reciprocal space maps of $N = 1$ to 5 superlattices taken around the (2 0 4) reflections of $SrTiO_3$ substrates. (**D**) The average and the ideal $c$-axis lattice constants of the superlattices. (**E**) The (3/2 1/2 L) half order diffraction peaks of the $N = 1, 3, 5$ superlattices. The reciprocal lattice units (r.l.u.) in (B), (C) and (E) are calculated using the $SrTiO_3$ substrate lattice. Error bars represent ±1 standard deviation.

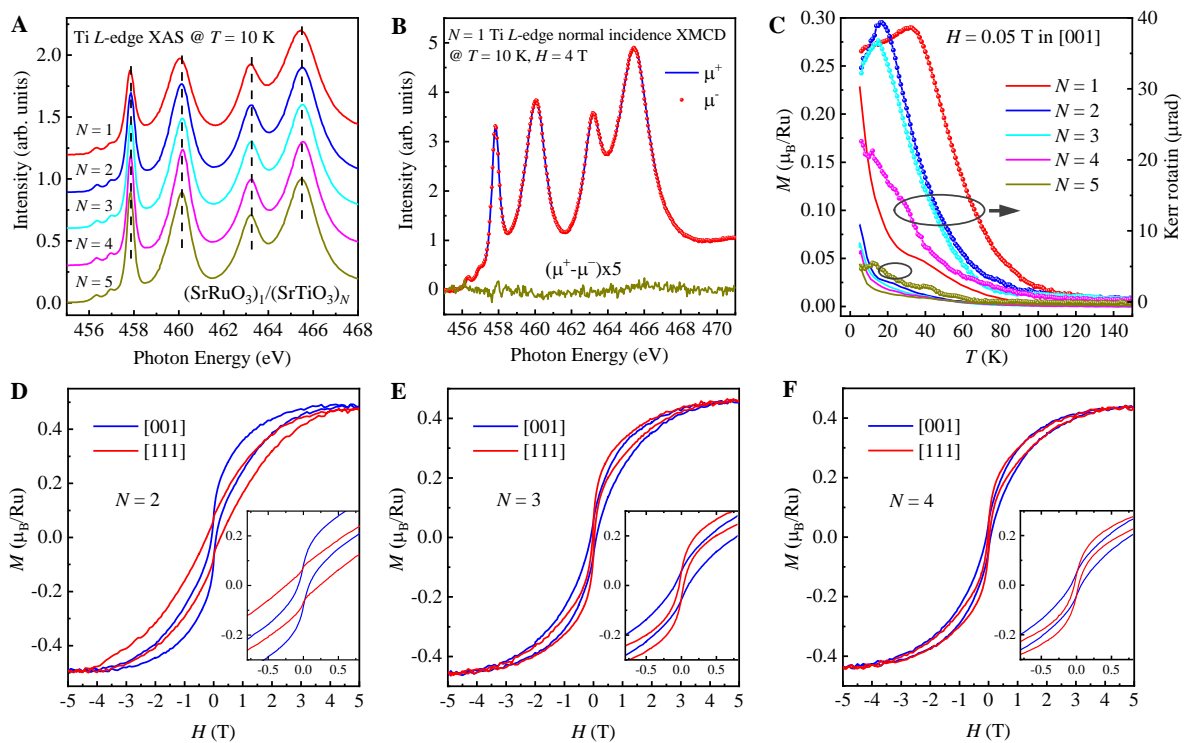

**Fig. 2. XMCD and SQUID magnetic characterizations of $(SrRuO_3)_1/(SrTiO_3)_N$ superlattices.** Ti $L$-edge (**A**) XAS and (**B**) XMCD of $N$ = 1-5 superlattices. (**C**) SQUID magnetization (left axis) and MOKE Kerr rotation (right axis) measurements as a function of temperature of $N$ = 1-5 superlattices. The measurements were taken during warming with 0.05 T field applied in the [001] direction. The magnetization versus magnetic field measured in the [001] and [111] directions of (**D**) $N$ = 2, (**E**) $N$ = 3 and (**F**) $N$ = 4 superlattices at 5 K. The insets are the zoom-in view of the loops at low field.

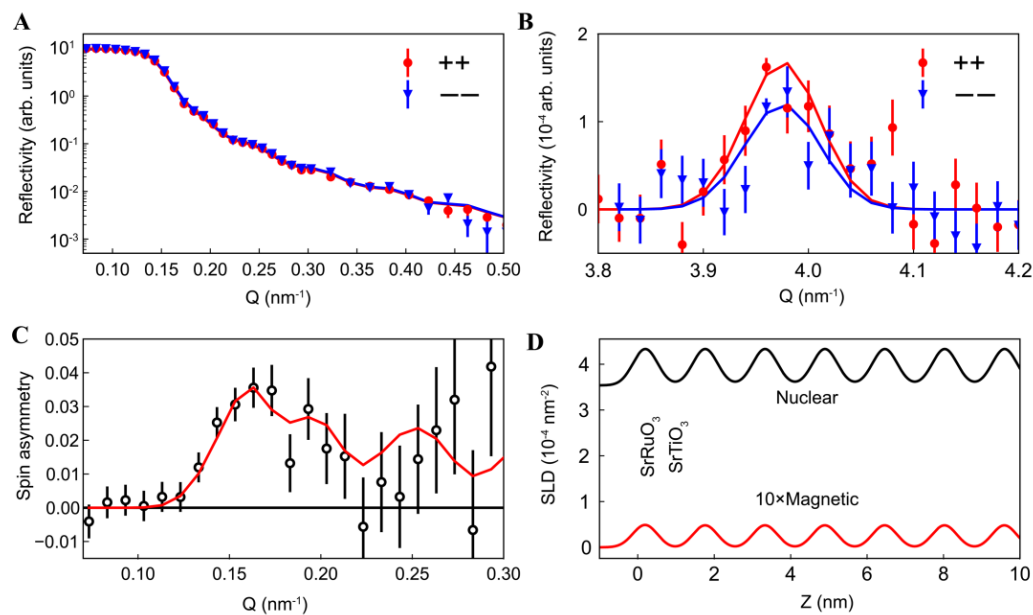

**Fig. 3. Polarized neutron reflectometry of a 50-repeat (SrRuO$_3$)$_1$/(SrTiO$_3$)$_3$ superlattice.** (**A**) Fitted PNR data. (**B**) Superlattice Bragg reflection fitted with Gaussian peaks to demonstrate the difference in peak height. (**C**) Spin asymmetry near the critical edge showing clear spin-dependent splitting of the reflectivities. All measurements were performed at 6 K under an applied field of 3 T. (**D**) Representative section of the nuclear and magnetic scattering length density (SLD) profiles used to generate the fits shown in (A) and (C). Error bars represent ±1 standard deviation.

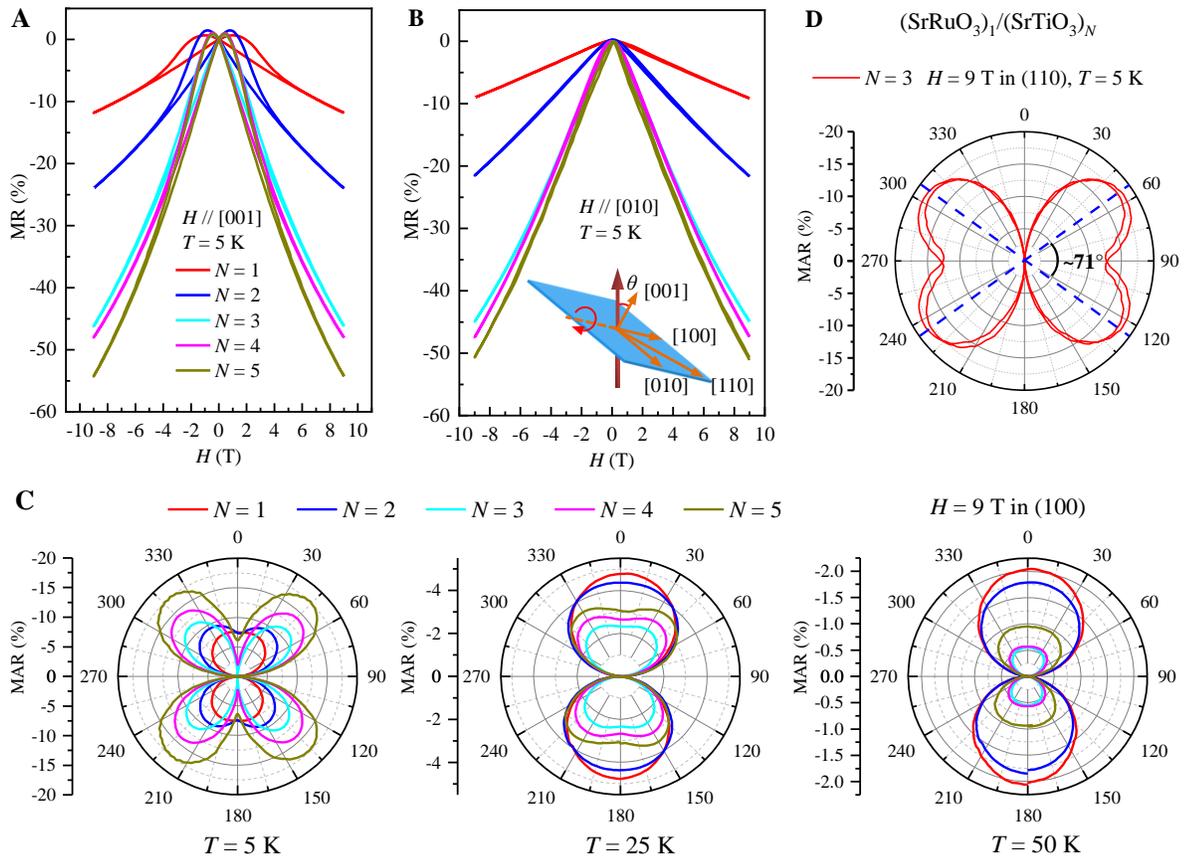

**Fig. 4. Magnetotransport properties of $(SrRuO_3)_1/(SrTiO_3)_N$ superlattices.** The MR at $T = 5$ K of $N = 1$-5 superlattices with the magnetic field applied parallel to (**A**) [001] and (**B**) [010] directions. The color correspondences are the same in (A) and (B). (**C**) Polar plots of MAR of $N = 1$-5 superlattices measured under a magnetic field of 9 T and at temperatures of 5 K, 25 K and 50 K. The geometry of the MAR measurement is shown in the inset of (B). The sample rotates around the [100] direction and the current is along the [100] direction, always being perpendicular to the magnetic field. $\theta$ is between [001] direction and the field direction within the (100) plane. (**D**) Polar plots of MAR of $N = 3$ superlattice measured under a magnetic field of 9 T and at temperature of 5 K. The sample rotates around the [110] direction and the current is along the [110] direction. $\theta$ is between [001] direction and the field direction within the (110) plane. Both MAR with the sample rotating clockwise and anticlockwise are shown.

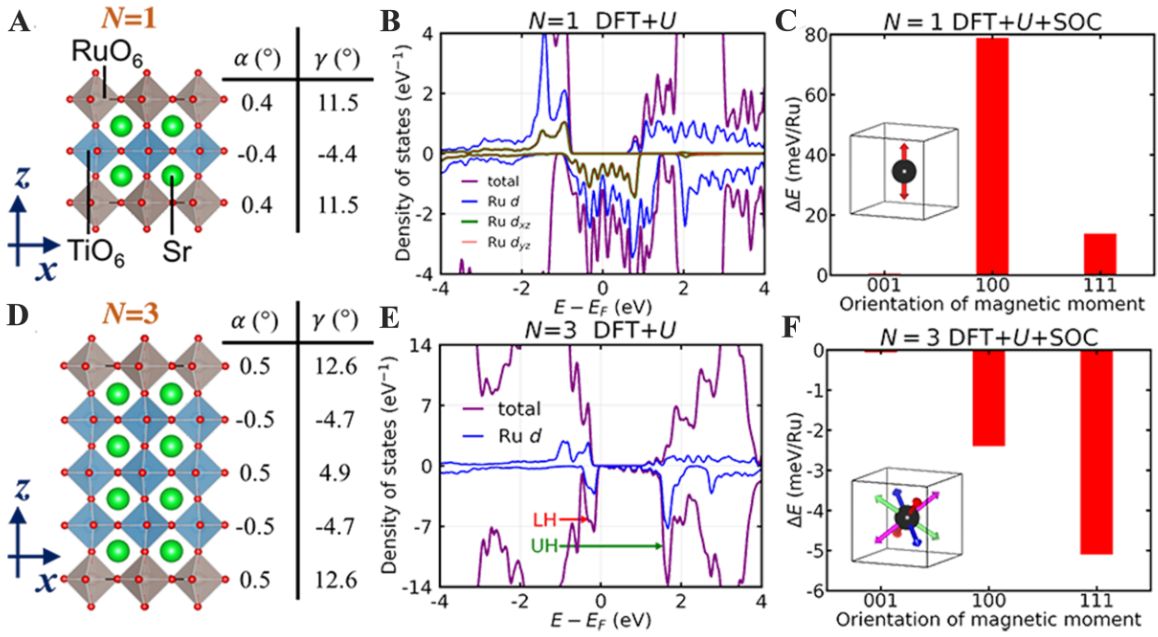

**Fig. 5. DFT calculated crystal structure, density of states and magnetic anisotropy of $(SrRuO_3)_1/(SrTiO_3)_N$ superlattices.** Crystal structures of (**A**) $N = 1$ and (**D**) $N = 3$ superlattices. Near-Fermi-level density of states of (**B**) $N = 1$ and (**E**) $N = 3$ superlattices, calculated using DFT+$U$ method with $U_{Ru} = 4$ eV. The states in the upper (lower) half correspond to spin up (down). In (E), 'LH' ('UH') means a lower (upper) Hubbard band, which is filled (empty). Due to the orbital ordering described in the main text, in each $RuO_2$ plane there are two distinct Ru atoms (labelled as Ru1 and Ru2): for Ru1, 'LH' is Ru (-) orbital and 'UH' is Ru (+); for Ru2, 'LH' is Ru (+) orbital and 'UH' is Ru (-) orbital. The definition of Ru (+) and Ru (-) orbitals can be found in the main text. Total energy of (**C**) $N = 1$ and (**F**) $N = 3$ superlattices with different magnetic moment orientations, calculated using DFT+$U$+SOC method with $U_{Ru} = 4$ eV. $\langle 001 \rangle$, $\langle 100 \rangle$ and $\langle 111 \rangle$ refer to the orientation of Ru magnetic moments. The energy of the $\langle 001 \rangle$ state is used as the reference. The twofold $\langle 001 \rangle$ MA is explicitly shown in the (C) inset. The eightfold $\langle 111 \rangle$ MA is explicitly shown in the (F) inset.

# Supplementary Materials for

## Correlation-driven eightfold magnetic anisotropy in a two-dimensional oxide monolayer

Zhangzhang Cui, Alexander J. Grutter, Hua Zhou, Hui Cao, Yongqi Dong, Dustin A. Gilbert, Jingyuan Wang, Yi-Sheng Liu, Jiaji Ma, Zhenpeng Hu, Jinghua Guo, Jing Xia, Brian J. Kirby, Padraic Shafer, Elke Arenholz, Hanghui Chen, Xiaofang Zhai, Yalin Lu

**This PDF file includes:**

- Fig. S1. Growth and structure characterizations of $(SrRuO_3)_1/(SrTiO_3)_N$ superlattices.

- Fig. S2. The Ru $K$-edge XANES spectra of $N = 1$, 3 and 5 superlattices.

- Fig. S3. Oxygen octahedral rotations along the (1 1/2 L) and (1/2 1/2 L) diffractions.

- Fig. S4. X-ray magnetic circular dichroism characterizations at 10 K.

- Fig. S5. Magnetic characterizations of $(SrRuO_3)_1/(SrTiO_3)_N$ superlattices.

- Fig. S6. Magnetic-field angle-dependent resistance characterizations and the determined magnetic easy axes.

- Fig. S7. Transport properties of $(SrRuO_3)_1/(SrTiO_3)_N$ superlattices.

- Fig. S8. DFT calculated crystal structure, density of states and magnetic anisotropy of $N = 5$ superlattice.

- Fig. S9. Calculated electronic structures of the $N = 1$ and $N = 3$ superlattices with and without oxygen octahedral tilts.

- Fig. S10. Orbital reconstruction of Ru $t_{2g}$ states.

- section SI. Qualitative analysis of the oxygen octahedral rotation pattern

- section SII. Role of oxygen octahedral tilts on electronic structure

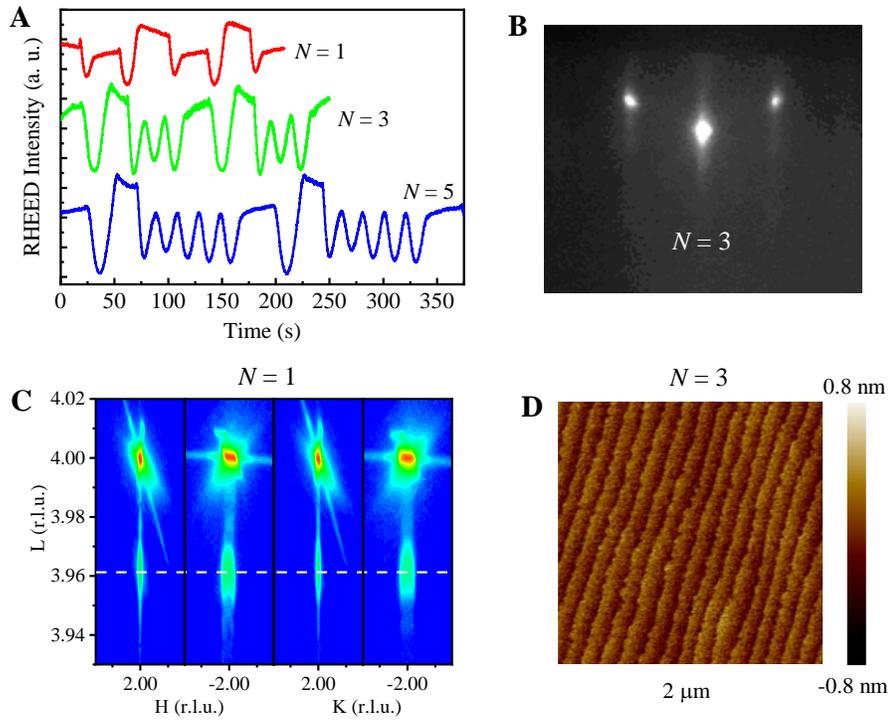

**Fig. S1. Growth and structure characterizations of (SrRuO$_3$)$_1$/(SrTiO$_3$)$_N$ superlattices.** (**A**) Reflective high energy electron diffraction (RHEED) intensity oscillation during the growth of $N$ =1, 3 and 5 superlattices. (**B**) RHEED diffraction pattern of $N$ = 3 superlattice. (**C**) X-ray diffraction (XRD) reciprocal space maps of $N$ = 1 superlattice taken around the (2 0 4), (-2 0 4), (0 2 4), and (0 -2 4) reflections of the SrTiO$_3$ substrate. The dash line indicates the diffraction from the superlattice. (**D**) The 2 μm × 2 μm surface atomic force microscopy image of the $N$ = 3 superlattice.

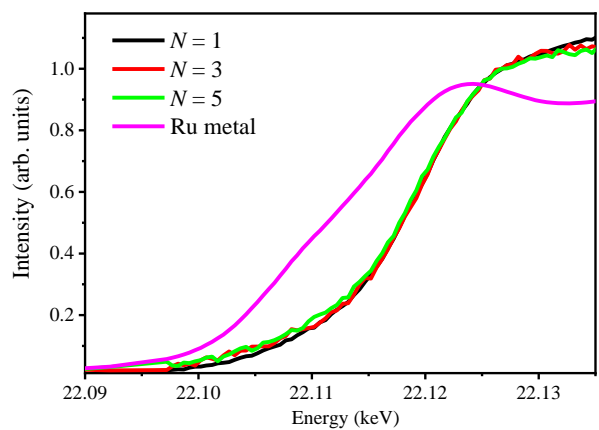

**Fig. S2. The Ru *K*-edge XANES spectra of *N* = 1, 3 and 5 superlattices.** The Ru metal foil is measured to calibrate the energy of the x-ray beam.

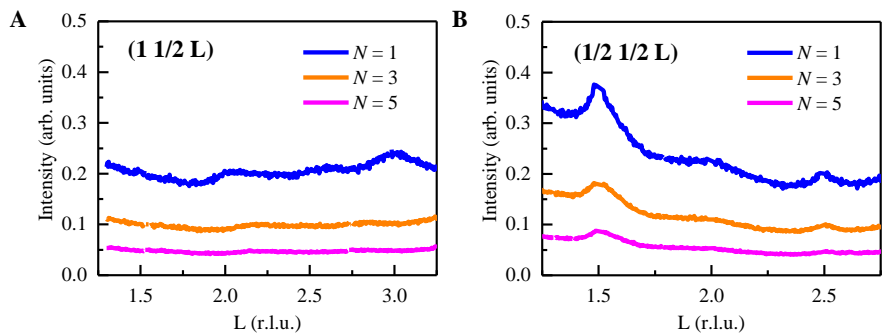

**Fig. S3. Oxygen octahedral rotations along the (1 1/2 L) and (1/2 1/2 L) diffractions.** XRD L scans along the (**A**) (1 1/2 L) and (**B**) (1/2 1/2 L) diffractions for the $N$ = 1, 3, and 5 superlattices.

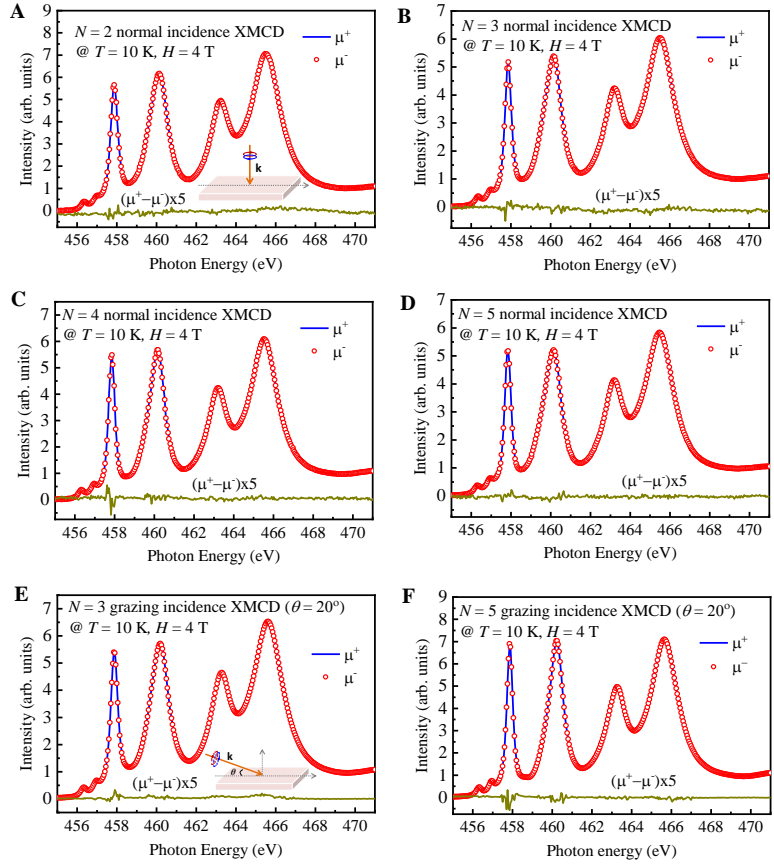

**Fig. S4. X-ray magnetic circular dichroism characterizations of (SrRuO$_3$)$_1$/(SrTiO$_3$)$_N$ superlattices at 10 K.** XMCD spectra with the incident beam perpendicular to the sample surface of (**A**) $N = 2$, (**B**) $N = 3$, (**C**) $N = 4$, and (**D**) $N = 5$ superlattices. XMCD spectra with the beam inclined with a grazing angle of 20° to the sample surface of (**A**) $N = 3$, (**B**) $N = 5$ superlattices. The above spectra were taken with left-polarized ($\mu^+$) and right-polarized ($\mu^-$) photons incident perpendicular or inclined to the sample surface.

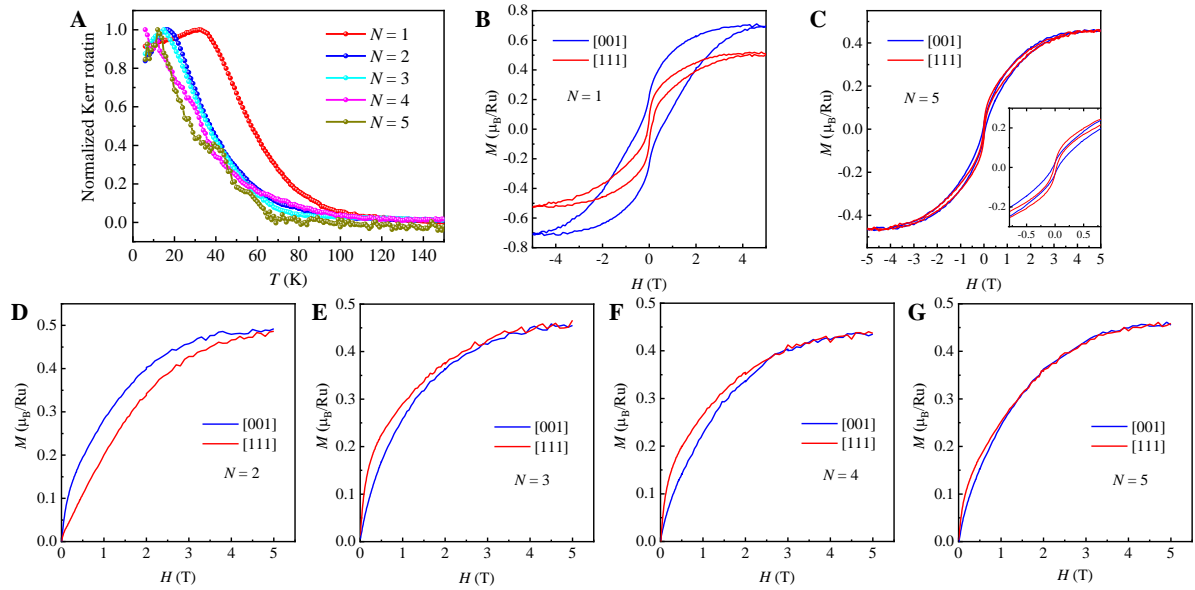

**Fig. S5. Magnetic characterizations of (SrRuO$_3$)$_1$/(SrTiO$_3$)$_N$ superlattices.** (**A**) Normalized Kerr rotation of $N$ = 1-5 superlattices. The magnetization hysteresis measurements with magnetic field along the [001] and [111] directions of (**B**) $N$ = 1 and (**C**) $N$ = 5 superlattices at 5 K. The inset shows the zoom-in view of the loops at low field. 0 T to 5 T magnetization measurements with magnetic field along the [001] and [111] directions of (**D**) $N$ = 2, (**E**) $N$ = 3, (**F**) $N$ = 4 and (**G**) $N$ = 5 superlattices.

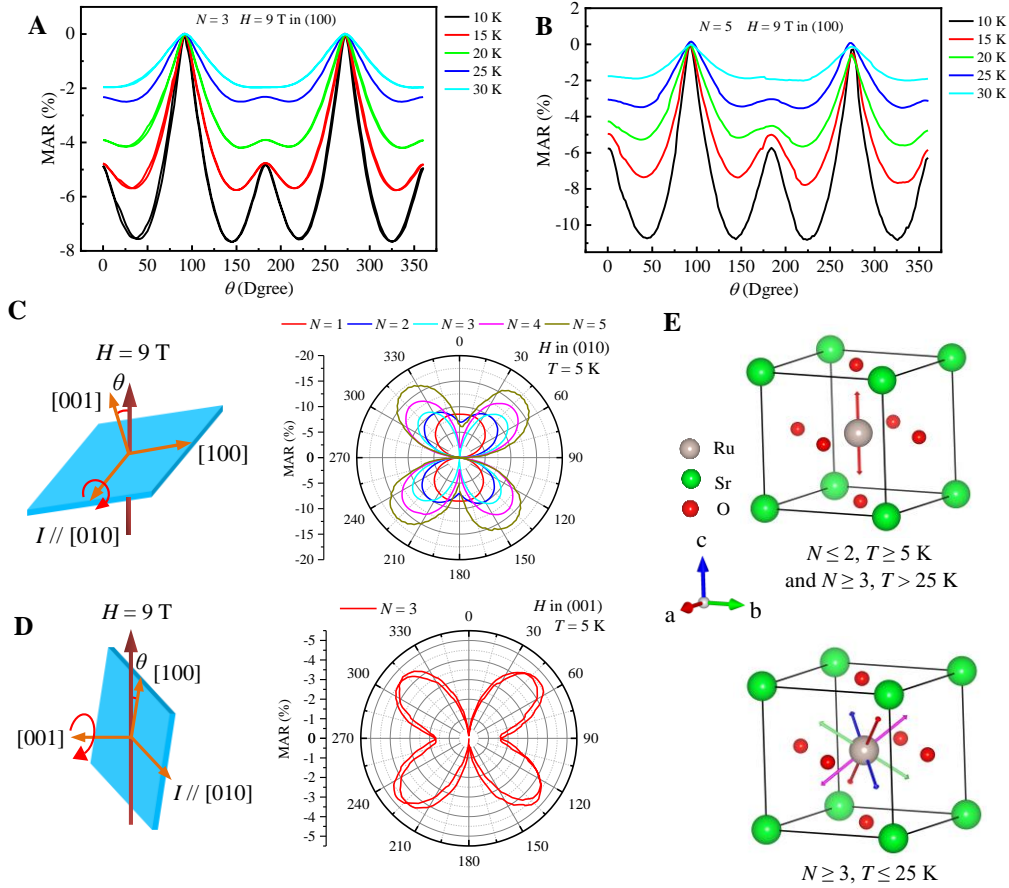

**Fig. S6. Magnetic-field angle-dependent resistance characterizations and the determined magnetic easy axes of (SrRuO$_3$)$_1$/(SrTiO$_3$)$_N$ superlattices.** Temperature dependences of MAR of (**A**) $N = 3$ and (**B**) $N = 5$ superlattices measured under a magnetic field of 9 T rotating in the (100) plane. At $T = 25$ K, there is noticeable fourfold symmetry overlapping on the twofold symmetry in the MAR. Therefore, the transition between uniaxial and eightfold magnetic anisotropy in $N \geq 3$ superlattices occurs at around 25 K. (**C**) Polar plots of the MAR of $N = 1$-5 superlattices measured under a magnetic field of 9 T and at 5 K with the magnetic field rotating in the (010) plane. (**D**) Polar plots of the MAR of $N = 3$ superlattice measured under a magnetic field of 9 T and at 5 K with the magnetic field rotating in the (001) plane. Both MAR with the magnetic field rotating clockwise and anticlockwise are shown. The schematics show the geometry in the transport measurements. (**E**) The determined magnetic easy axes of the SrRuO$_3$ lattice, shown by the arrows. The $N \leq 2$ superlattices at $T \geq 5$ K and $N \geq 3$ superlattices at $T > 25$ K have uniaxial MA and the magnetic easy axis is along the $\langle 001 \rangle$ direction of the SrRuO$_3$ lattice. The $N \geq 3$ superlattices at $T \leq 25$ K have eightfold MA and the magnetic easy axes are along the $\langle 111 \rangle$ directions of the SrRuO$_3$ lattice.

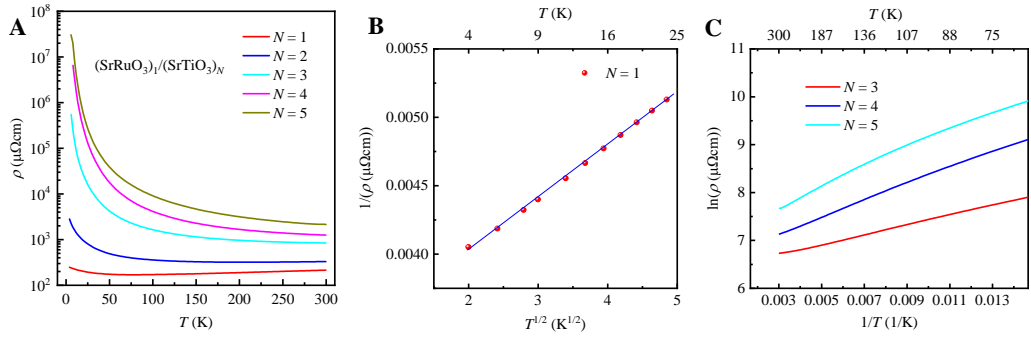

**Fig. S7. Transport properties of (SrRuO₃)₁/(SrTiO₃)ₙ superlattices.** (**A**) Temperature-dependent resistivity ($\rho$) of $N = 1$-5 superlattices. (**B**) The resistivity fitting of $N = 1$ superlattice at low temperatures (4 K – 25 K). The resistivity is best fitted to the electron-electron correlation induced localization model ($1/\rho \propto -T^{1/2}$) (P. A. Lee, *Rev. Mod. Phys.* **57**, 287-337 (1985)). (**C**) The relation between logarithmic resistivity and the reciprocal temperature of $N = 3$, 4 and 5 superlattices at relatively higher temperatures (70 K – 300 K). From the slopes of the nearly linear relationship, the band gaps (twice the thermal activation energy) are estimated to be 18.0 meV ($N = 3$), 30.2 meV ($N = 4$) and 34.0 meV ($N = 5$) respectively.

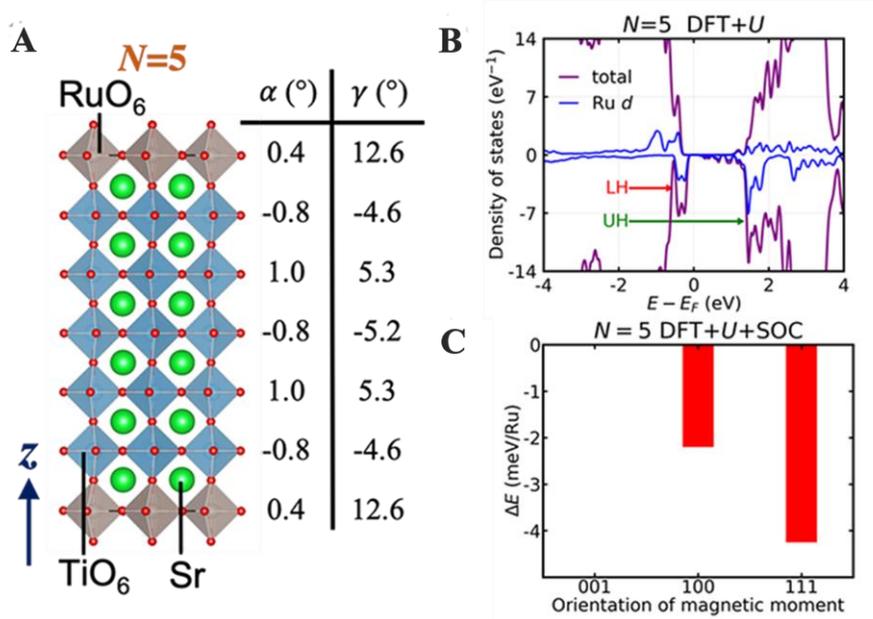

**Fig. S8. DFT calculated crystal structure, density of states and magnetic anisotropy of $N = 5$ superlattice.** (**A**) Crystal structure of $N = 5$ superlattice. (**B**) Near-Fermi-level density of states of $N = 5$ superlattice, calculated using DFT+$U$ method with $U_{Ru} = 4$ eV. The $N = 5$ superlattice shows a semiconducting density of states. The purple lines are total density of states and the blue lines are Ru $d$ projected density of states. 'LH' ('UH') means a lower (upper) Hubbard band, which is filled (empty). Due to the orbital ordering described in the main text, in each RuO$_2$ plane there are two distinct Ru atoms (labelled as Ru1 and Ru2): for Ru1, 'LH' is Ru (−) orbital and 'UH' is Ru (+) orbital; for Ru2, 'LH' is Ru (+) orbital and 'UH' is Ru (−) orbital. The definition of Ru (+) and Ru (−) orbitals can be found in the main text. (**C**) Total energy of $N = 5$ superlattice with $\langle 001 \rangle$, $\langle 100 \rangle$ and $\langle 111 \rangle$ magnetic moment orientations, calculated using DFT+$U$+SOC method with $U_{Ru} = 4$ eV. The energy of the $\langle 001 \rangle$ state is used as the reference.

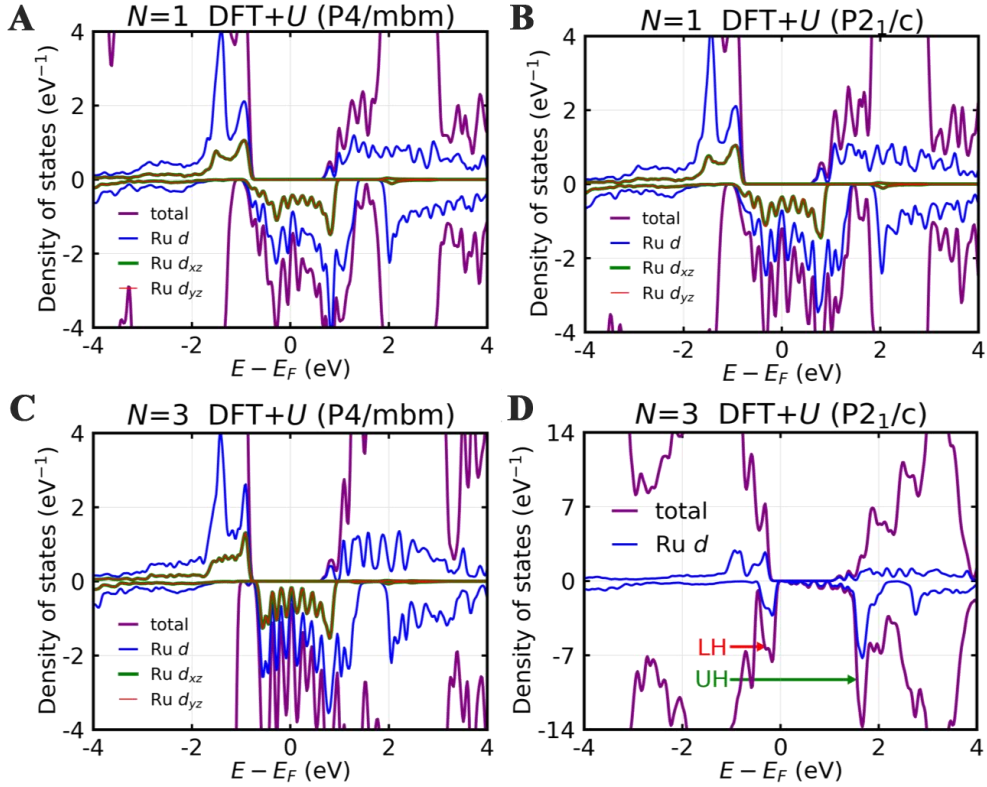

**Fig. S9. Calculated electronic structures of the $N = 1$ and $N = 3$ superlattices with and without oxygen octahedral tilts.** (**A**) Densities of states (DOS) of the $N = 1$ superlattice without oxygen octahedral tilts; The structure has P4/mbm symmetry (No. 127). (**B**) DOS of the $N = 1$ superlattice with oxygen octahedral tilts; The structure has P2$_1$/c symmetry (No. 14). (**C**) DOS of the $N = 3$ superlattice without oxygen octahedral tilts; The structure has P4/mbm symmetry (No. 127). (**D**) DOS of the $N = 3$ superlattice with oxygen octahedral tilts; The structure has P2$_1$/c symmetry (No. 14). All DOS are calculated using density functional theory (DFT)+$U$ method with $U_{Ru} = 4$ eV. The states in the upper (lower) half correspond to spin up (down). The purple and blue curves are total density of states and partial density of states projected onto Ru $d$ orbitals. In (A), (B) and (C), the green and red curves are partial densities of states projected onto Ru $d_{xz}$ and Ru $d_{yz}$ orbitals. In panel (D), the red arrow refers to a lower Hubbard band ("LH", which is filled) and the green arrow refers to an upper Hubbard band ("UP", which is empty). Due to the orbital ordering described in the text, in each RuO$_2$ plane there are two distinct Ru atoms (labelled as Ru1 and Ru2): for Ru1, "LH" is Ru (-) orbital and "UH" is Ru (+); for Ru2, "LH" is Ru (+) orbital and "UH" is Ru (-) orbital. The definition of Ru (+) and Ru (-) orbitals can be found in the main text.

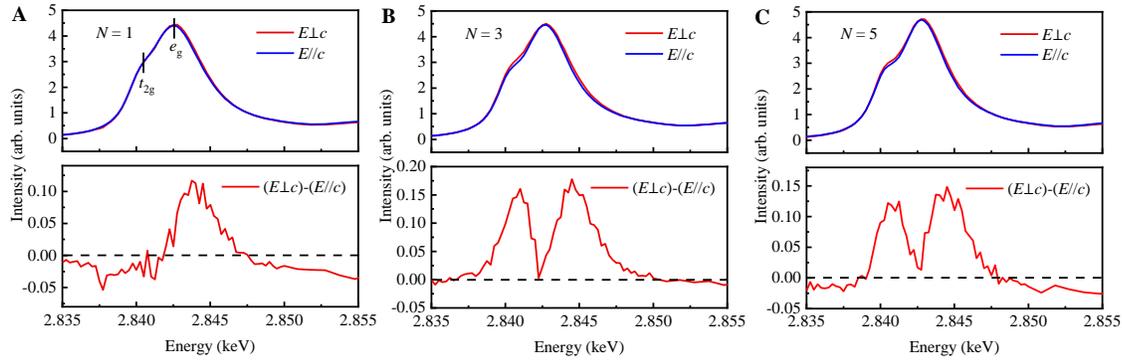

**Fig. S10. Orbital reconstruction of Ru $t_{2g}$ states.** X-ray linear dichroism (XLD) spectra of (**A**) $N = 1$, (**B**) $N = 3$, and (**C**) $N = 5$ superlattices. The spectra were measured at Ru $L_3$-edge with the incident photon polarization perpendicular ($E \perp c$) and parallel ($E // c$) to the superlattice $c$ axis respectively. The two absorption peaks around 2.842 keV (indicated by solid lines) are due to the transition from Ru $2p_{3/2}$ to Ru $4d$ $t_{2g}$ and $e_g$ orbitals respectively. At $E \perp c$, the Ru $t_{2g}$ absorption is dominated by Ru $4d_{xy}$ and Ru $4d_{yz,zx}$ transitions. Whereas, at $E // c$, it is dominated by Ru $4d_{yz,zx}$ transition. It can be clearly seen that in $N = 1$ superlattice, there is very weak XLD signal of the Ru $t_{2g}$ orbitals, indicating no obvious orbital polarization between Ru $4d_{xy}$ and Ru $4d_{yz,zx}$ states in $N = 1$ superlattice. Whereas, in $N = 3$ and 5 superlattices, the Ru $t_{2g}$ orbitals show a prominent XLD peak. It suggests that the occupations in Ru $4d_{xy}/4d_{xz}$ and $4d_{yz}$ orbitals of $N = 3$ and 5 superlattices are very different from those of $N = 1$ superlattice while XLD signals of $N = 3$ and $N = 5$ superlattices are rather similar. Therefore, the XLD results provide evidence of redistribution of Ru orbital occupancies from $N = 1$ to $N = 3$ and $N = 5$ superlattices. This experimental observation and the trend are also in line with the magnetic anisotropy transition from $N = 1$ to $N = 3, 5$ superlattices.

**section SI. Qualitative analysis of the oxygen octahedral rotation pattern**

We observed no diffraction intensities for peak indices with one integer (*eg.* 1/2, 1, 3/2; etc.) or two integers (*eg.* 1/2, 1, 2; etc.). Thus the possibilities of in-phase (+) rotations or A-site cation displacements are excluded which indicates the rhombohedral type $a^-b^-c^-$ rotation (S. May, *Phys. Rev. B* **82**, 014110 (2010); S. May, *Phys. Rev. B* **82**, 014110 (2010)). Since the SrTiO$_3$ substrate has the square symmetry in the in-plane direction, previous studies have found that the epitaxial films have the same rotational angles around the *a* and *b* directions (S. May, *Phys. Rev. B* **82**, 014110 (2010); S. May, *Phys. Rev. B* **82**, 014110 (2010); X. Zhai, *Nat. Commun.* **5**, 4283 (2014)). Thus the $a^-b^-c^-$ rotation is simplified to the $a^-a^-c^-$ rotation.

In the current study, since the half-order Bragg peaks are from the oxygen atoms that are displaced from their ideal positions due to oxygen octahedral rotations, the peak intensity at (*H*, *K*, *L*) can be written as

$$I(H,K,L) \propto \sum_{j=1}^{4} D_j \left| f_{O^{2-}} \exp\left(-\frac{B(H^2+K^2+L^2)}{4a^2}\right) \sum_{n=1}^{24} \exp[2\pi i(Hu_n + Kv_n + Lw_n)] \right|^2, \quad (1)$$

where $D_j$ is the domain occupancy, $f_{O^{2-}}$ is the x-ray form factor for O$^{2-}$ [3], $B$ is the thermal broadening (Debye-Waller) factor, $a$ is the pseudocubic lattice constant, and ($u_n$, $v_n$, $w_n$) are the positions of *n*th oxygen atom in units of the real space lattice (X. Zhai, *Nat. Commun.* **5**, 4283 (2014)). The pseudocubic lattice constant is taken to be the same along all three axes, as the small difference along the *c*-axis is a higher order effect and can be ignored. The octahedral rotation effectively doubles the unit cells along each of the pseudocubic axis, resulting in a supercell with 24 oxygen atoms. The positions of the oxygen atoms can be individually calculated from the rotation matrix and their scattering amplitudes are added coherently within each domain according to eq. (1).

For the $a^-a^-c^-$ rotation system, the half-order peak intensity of the first domain with rotation angle of (*α*, *α*, *γ*) can be analytically written as

$$I(H,K,L) \propto D_j \sum_j \left\{ f_{O^{2-}}(\sqrt{H^2+K^2+L^2}) \exp\left(-\frac{B(H^2+K^2+L^2)}{4a^2}\right) \zeta \right\}^2, \quad (2)$$

$$\zeta = \begin{cases} 0 & \text{if any of H,K,L is integer} \\ [\sin\pi(K-H)\alpha \sin\pi L + \sin\pi(L\alpha - K\gamma)\sin\pi H + \sin\pi(H\gamma - L\alpha)\sin\pi K] & \text{otherwise} \end{cases}$$

where *γ* is the rotation angle about the *c*-axis and *α* is the rotation angle about the *a*-axis or *b*-axis. The other three domains have different sense of rotation angles of (-*α*, *α*, *γ*), (*α*, -*α*, *γ*) and (*α*, *α*, -*γ*). Their corresponding diffraction equations can be written by simply alternating the sign of *H* or *K* or *L* in equation (2). For example, to obtain the diffraction equation for the second domain with rotation angles of (-*α*, *α*, *γ*), it is only needed to change the sign of *H* in equation (2). Then the overall diffraction intensity from all four domains can be calculated. Usually for epitaxial films on square lattices with the fourfold symmetry, the occupations of four domains are equal. Thus by assuming an equal occupation of four domains and both *α* and *γ* being small, it is found that the overall diffraction intensities for *H* = *K* peaks are roughly proportional to $\alpha^2$. While the overall diffraction intensities for *H* ≠ *K* peaks are dependent on both $\alpha^2$ and $\gamma^2$. By comparing the intensities of the *H* ≠ *K* peaks shown in Fig. 1E and the *H* = *K* peaks shown in Fig. S3B, it is found the former peaks are one order or two orders of magnitude larger than the latter peaks. Therefore the rotation angle *α* is much smaller than *γ* in all three superlattices.

**section SII. Role of oxygen octahedral tilts on electronic structure**

In Supplementary Fig. S9, we compare the densities of states (DOS) of the $N = 1$ and $N = 3$ superlattices in two different structures. One structure has P4/mbm symmetry (space group No. 127) with only in-plane rotations of oxygen octahedra but no out-of-plane rotations (i.e. no oxygen octahedral tilts). The other structure has P2$_1$/c symmetry (space group No. 14) with both in-plane rotations and out-of-plane rotations of oxygen octahedra. All the DOS are calculated by DFT + U method with $U_{Ru} = 4$ eV. Panel (A) shows the DOS of the $N = 1$ superlattice with the P4/mbm structure, demonstrating that the superlattice is conductive with Ru $d_{xz}$ and Ru $d_{yz}$ orbitals being degenerate. Panel (B) shows the DOS of the $N = 1$ superlattice with the P2$_1$/c structure, which is essentially the same as that of the $N = 1$ superlattice with the P4/mbm structure. This indicates that the small tilts of oxygen octahedral do not change the electronic structure of the $N = 1$ superlattice, which is corroborated with the fact that the total energy of the P2$_1$/c structure is lower than that of the P4/mbm structure by only 0.1 meV/Ru. Panel (C) shows the DOS of the $N = 3$ superlattice with the P4/mbm structure, demonstrating that the superlattice is conductive with Ru $d_{xz}$ and Ru $d_{yz}$ orbitals being degenerate. This electronic structure is similar to that of the $N = 1$ superlattice with the P4/mbm structure, with the band width of Ru $d_{xz}$ and Ru $d_{yz}$ orbitals slightly reduced in the $N = 3$ superlattice. However, with the small tilts of oxygen octahedra, the electronic structure of the $N = 3$ superlattice becomes fundamentally different. Panel (D) shows the DOS of the $N = 3$ superlattice with the P2$_1$/c structure, demonstrating that the superlattice is semiconducting with a small gap (about 0.1 eV). The Ru $d_{xz}$ and Ru $d_{yz}$ orbitals hybridize with each other and form a pair of new orbitals Ru $\alpha|xz\rangle + \beta|yz\rangle$ orbital (referred to as Ru (+) state) and Ru $\beta|xz\rangle - \alpha|yz\rangle$ orbital (referred to as Ru (-) state). In each RuO$_2$ plane, there are two distinct Ru atoms: on one Ru atom, the fourth electron fills Ru (+) state and leaves Ru (-) state empty; on the other Ru atom, the fourth electron fills Ru (-) state and leaves Ru (+) state empty. The filled new orbital is referred to as a lower Hubbard band, which is just below the Fermi level. Since a new orbital ordering emerges with the tilts of oxygen octahedral, the total energy of the $N = 3$ superlattice with the P2$_1$/c structure is significantly lower than that of the P4/mbm structure by 146 meV/Ru. In summary, Fig. S9 shows that in order to induce the new orbital ordering and the resulting eightfold magnetic anisotropy in SrRuO$_3$ monolayers, we need an enhanced correlation strength on Ru ions by increasing *N* and tilts of oxygen octahedra that lower the crystal symmetry.